\documentclass[osajnl,twocolumn,showpacs,superscriptaddress,10pt]{revtex4-1} 
\usepackage[latin1]{inputenc}
\usepackage{amsfonts}
\usepackage{mathptmx}
\usepackage{amsmath,amssymb,graphicx}
\usepackage{epstopdf}
\begin{document}

\title{Mid-Infrared Soliton and Raman Frequency Comb Generation in Silicon Microrings}
\author{Tobias Hansson}
\affiliation{Dipartimento di Ingegneria dell'Informazione, Universit\`a di Brescia, via Branze 38, 25123 Brescia, Italy}
\affiliation{Department of Applied Physics, Chalmers University of Technology,  SE-41296 G\"oteborg, Sweden}
\author{Daniele Modotto}
\author{Stefan Wabnitz}\email{Corresponding author: stefan.wabnitz@unibs.it}
\affiliation{Dipartimento di Ingegneria dell'Informazione, Universit\`a di Brescia, via Branze 38, 25123 Brescia, Italy}


\begin{abstract}
We numerically study the mechanisms of frequency comb generation in the mid-infrared spectral region from cw pumped silicon microring resonators. Coherent soliton comb generation may be obtained even for a pump with zero linear cavity detuning, through suitable control of the effective lifetime of free-carriers from multiphoton absorption, which introduces a nonlinear cavity detuning via free-carrier dispersion. Conditions for optimal octave spanning Raman comb generation are also described.   
\end{abstract}


\maketitle

\baselineskip 11pt

Optical frequency comb generation using nonlinear optical microresonators offers an intriguing alternative to mode-locked lasers. A host of potential applications, including biomedical and environmental spectroscopy, arise in the mid-infrared (MIR) range of the spectrum \cite{Wang13}. Planar microring resonators that can be fabricated directly on CMOS compatible chips are particularly interesting for their low cost and mass manufacturing potential. The natural candidate for implementing a chip-scale, MIR frequency comb source is silicon, because of its $10^2$ higher nonlinearity compared to silica, coupled with vanishing two-photon absorption (TPA) and associated free-carrier absorption (FCA) and dispersion (FCD) for photon energies below the silicon half band-gap. Recent experiments have demonstrated that silicon-chip based MIR frequency comb generation is critically dependent upon the possibility of reducing the lifetime of free carriers generated by three-photon absorption (3PA) \cite{griff14a,griff14b}.

In this work we present a numerical study of MIR frequency comb generation in silicon microresonators, which demonstrates that a proper control of the free-carrier lifetime (FCT) may enable a new route for stable soliton self-mode-locking \cite{saha13,Herr14} of the comb. Indeed, multiphoton absorption induced FCD introduces a dynamic nonlinear cavity detuning, which replaces the need for a nonzero linear cavity detuning \cite{Herr14,Lamo13}. We also predict the highly efficient generation of MIR Raman frequency combs \cite{Couny07} from silicon microresonators.
           
The temporal dynamics of frequency comb generation in silicon microresonators is described by a generalized nonlinear envelope equation (GNEE) for the field envelope $A\:[\sqrt{W}/m]$ which includes linear loss and dispersion, the Kerr effect, Raman scattering, TPA and 3PA, FCA and FCD \cite{Lamo13}-\cite{HMW13a}
\begin{align}
& \left[\frac{1}{v_g}\partial_{\tau} - D + \alpha +i\delta_0+\frac{\sigma}{2}\left(1+i\mu\right)\left\langle N_c(\tau)\right\rangle\right]A(\tau,t)= \nonumber\\
& f_0+ik_0\left(1+i\tau_{sh}\frac{\partial}{\partial t}\right)p_{NL}^{(3)}(\tau,t)
\label{GNEE}
\end{align}
\noindent coupled with the evolution equation for the averaged carrier density $\left\langle N_c(\tau)\right\rangle$
\begin{equation}
 \partial_{\tau}\left\langle N_c(\tau)\right\rangle = \frac{\beta_{TPA}(\omega)}{2\hbar\omega_0}\left\langle |A|^4\right\rangle+\frac{\gamma_{3PA}}{3\hbar\omega_0}\left\langle |A|^6\right\rangle-\frac{\left\langle N_c(\tau)\right\rangle}{\tau_{eff}}
\label{Carrier}
\end{equation}
where $\tau$ is a continuous (slow) temporal variable that replaces the round-trip number, $v_g=c/n_g$ and $n_g$ are the group velocity and index at the pump carrier frequency $\omega_0$, $k_0=\omega_0/c$, $\tau_{sh}=1/\omega_0$, and $t$ is a retarded (fast) time. In Eq.(\ref{Carrier}), brackets denote average over the cavity circulation time $t_R$: $\left\langle X(\tau)\right\rangle = (1/t_R)\int_{-t_r/2}^{t_r/2} X(t,\tau) dt$, so that Eq.(\ref{Carrier}) describes the buildup of carriers within the cavity over many round trips, supposing that the FCT $\tau_{eff}\gg t_R$. The group-velocity dispersion (GVD)
operator $D$ reads as
\begin{equation}
 D = \sum_{m\geq2}\frac{i^{m+1}}{m!}\left(\frac{\partial^m \beta}{\partial \omega^m}\right) \frac{\partial^m}{\partial t^m}.
\label{GVD}
\end{equation}
\noindent In Eq.(\ref{GNEE}) the linear loss coefficient $\alpha =\alpha'/L$ with $\alpha'= \alpha_dL+\theta/2$, $\alpha_d$ represents distributed cavity loss and $\theta$ is the transmission coefficient between the resonator of length $L$ and the bus waveguide. Moreover $\sigma$ is the FCA coefficient, $\mu$ is the FCD coefficient, $\beta_{TPA}$ is the TPA coefficient,  $\delta_0=\bar{\delta_0}/L=(\omega_R-\omega_0)t_R/L$ is the linear cavity detuning (where $\omega_R$ is the closest linear cavity resonance to the pump frequency $\omega_0$), $f_0=(\sqrt{\theta}/L)A_{in}$ and $A_{in}$ the injected cw pump amplitude. The cavity boundary conditions impose the field to be $t-$periodic with period $t_R$, i.e., $A(\tau,t)= A(\tau,t+t_R)$. The nonlinear polarization reads as 
\begin{align}
& p_{NL}^{(3)}=n_2[\left(1-\gamma_R\right)(\eta \otimes (|A|^2A)+ir_{3}|A|^4A\nonumber\\
& +A^3\exp(-2i\omega_0t)/3)+\gamma_R\int_{-\infty}^t h_R(t-t')|A(t')|^2dt'] 
\label{p3}
\end{align}
\noindent where $\otimes$ denotes convolution product, $n_2$ is the nonlinear index and $\gamma_R$ is the Raman fraction coefficient associated with the response function 
\begin{equation}
 h_R(t) = H(t)\frac{\tau_1^2+\tau_2^2}{\tau_1\tau_2^2}\exp(-t/\tau_2)\sin(t/\tau_1),
\label{raman}
\end{equation}
\noindent where $H(t)$ is the Heaviside function, $\tau_1=10.2\:fs$ and $\tau_2=3.03\:ps$. Whenever pumping the transverse electric (TE) mode, the scalar Raman gain coefficient $\gamma_R=g_R\Gamma_R/(n_2k_0\Omega_R)=0.018$, where we used $n_2=9\times10^{-18}m^2/W$ and the peak parallel Raman gain value $g_R=2\times10^{-10}m/W$ at the wavelength $\lambda_0=1.55\:\mu m$, the gain bandwidth $\Gamma_R/\pi=105\: GHz$ and the peak gain frequency shift $\Omega_R/2\pi=15.6\: THz$. On the other hand, when the input cw pump is coupled to the transverse magnetic (TM) mode, the parallel Raman gain vanishes in silicon \cite{LPA07}, hence $\gamma_R=0$. 
%
%
In Eq.(\ref{p3}) $\eta=\delta(t)+r_2(t)$, where $r_2(t)=\mathcal{F}^{-1}(\hat{r_2}(\omega))$ is related to the frequency dependent TPA coefficient as $\hat{r_2}(\omega) = \beta_{TPA}(\omega)c/(2\omega_0 n_2)$, where the wavelength dependence of the TPA coefficient $\beta_{TPA}(\omega)$ was estimated using the analytical approximation of Ref.\cite{BRD07}: TPA vanishes for wavelengths larger than $2.2\:\mu m$. Moreover, we supposed a wavelength independent 3PA contribution which is represented in Eq.(\ref{p3}) by the coefficient $r_3=\gamma_{3PA}c/3\omega_0 n_2$.
\begin{figure}[htbp]
\centerline{\includegraphics[width=0.8\columnwidth]{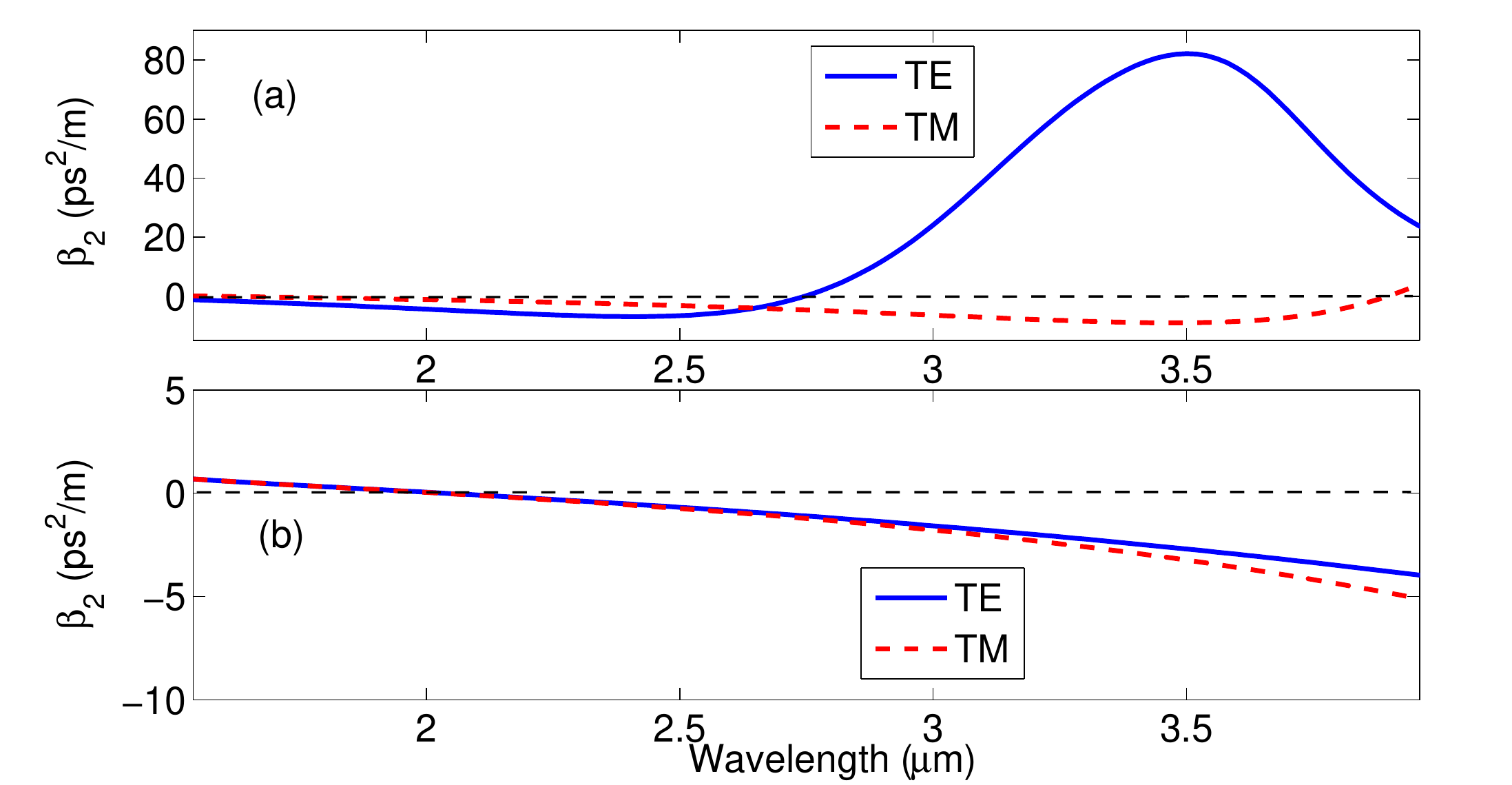}}
  \caption{GVD $\beta_2$ for the TE (solid blue) and TM (dashed red) modes of a silicon ring resonator with (a) inner radius a = 49.5 $\mu m$, outer radius b = 50.5 $\mu m$, and h = 0.5 $\mu m$ (see Ref\cite{HMW14}) or (b) for waveguide dimensions of 1.5x1.5 $\mu m^2$}
\label{fig:fig2}
\end{figure}

We tailored the linear dispersion and effective area properties of the silicon microrings for efficient MIR frequency comb generation. This involves a suitable design of the microresonator radius and cross section. In fact, the shape of the frequency combs can be controlled through the geometric contribution to the effective refractive index of the cavity mode. Fig.\ref{fig:fig2}(a) illustrates examples of dispersion profiles for the TE and TM modes of a 50 $\mu m$ radius ring, with a 1 $\mu m$ wide, 500 nm thick waveguide, computed from a commercial mode solver (for details, see Ref.\cite{HMW14}). Whereas in Fig.\ref{fig:fig2}(b) we show the dispersion profiles for a 1.5 $\mu m$ wide, 1.5 $\mu m$ thick waveguide: as can be seen, the absolute value of dispersion is considerably reduced with respect to the previous case of a smaller waveguide, in particular as far as the TE mode is concerned. Moreover, for the larger waveguide the GVD of both modes remains anomalous for $\lambda> 2 \mu m$.
%
%
The effective area of the TM mode of the large (small) waveguide is equal to $1.37\:\mu m^2 $ ($0.32\:\mu m^2 $) at $\lambda_0=2.6\:\mu m$; whereas for the TE modes $A_{eff}=1.23\:\mu m^2 $ ($A_{eff}=0.32\:\mu m^2 $). 
%
\begin{figure}[htbp]
\centerline{\includegraphics[width=0.8\columnwidth]{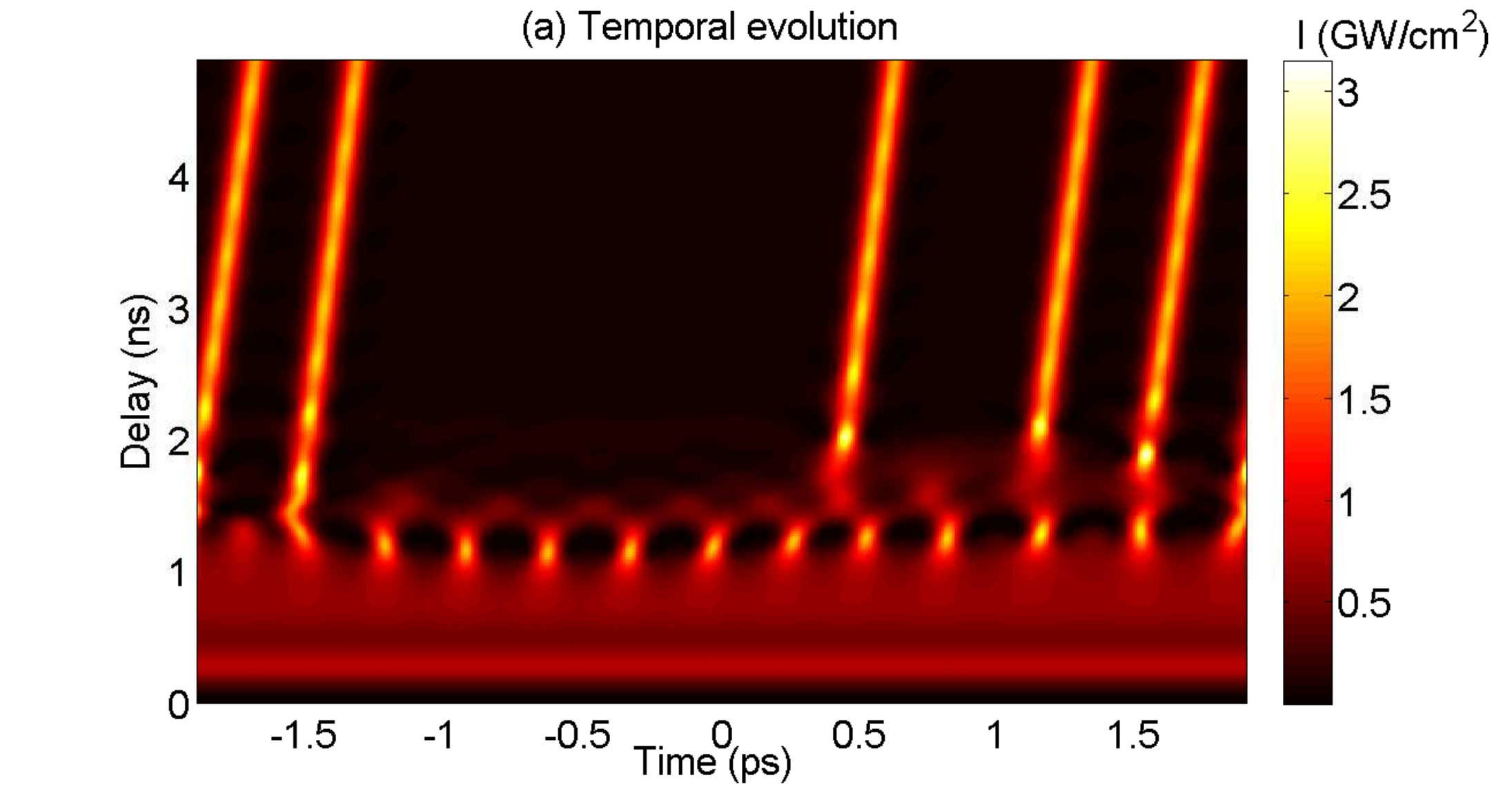}}
\centerline{\includegraphics[width=0.8\columnwidth]{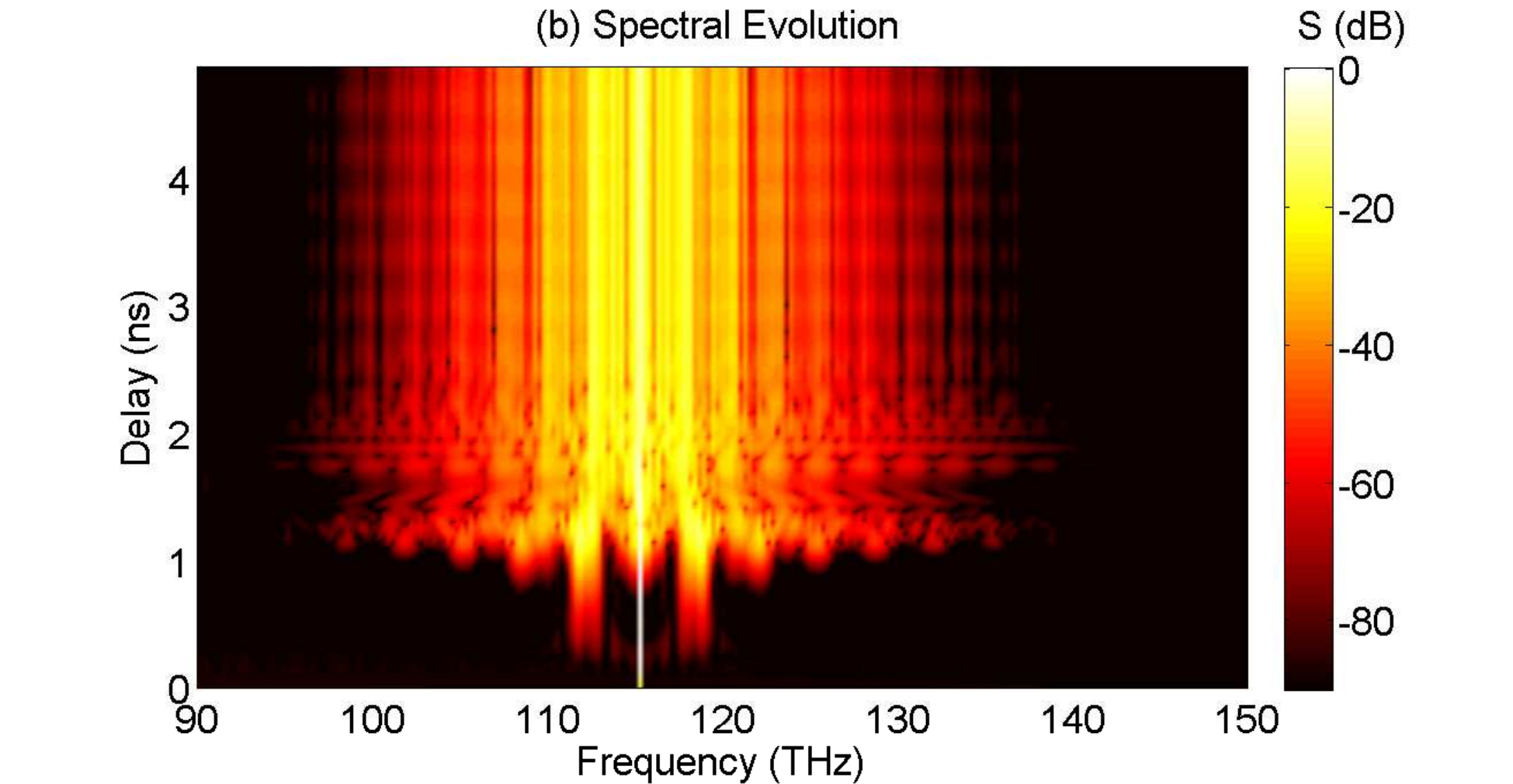}}
\centerline{\includegraphics[width=0.8\columnwidth]{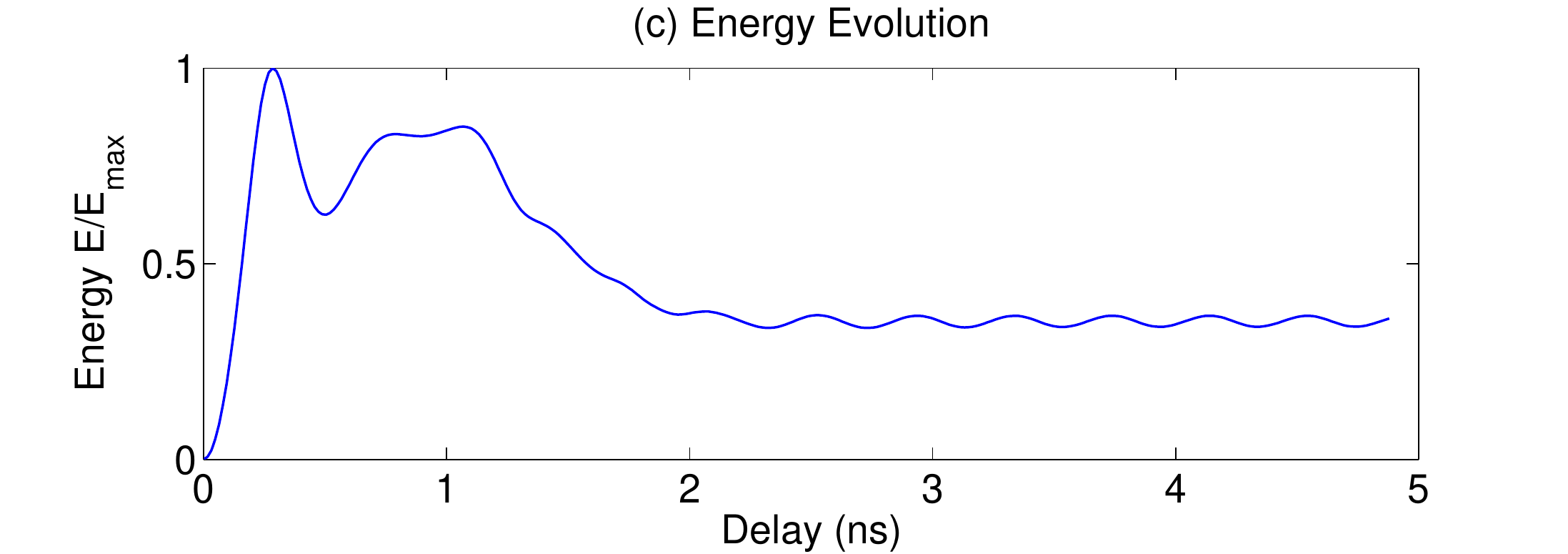}}
  \caption{Soliton frequency comb generation from TM mode of large microring as in Fig.\ref{fig:fig2}(b): (a) temporal evolution; (b) spectral evolution of intracavity intensity; (c) energy evolution. Here $\tau_{eff}=320\:ps$ and $\lambda_0=2.6\:\mu m$}
\label{fig:fig4}
\end{figure}

Based on our design of the spatial mode properties of the microrings, we numerically simulated the temporal dynamics of MIR frequency comb generation by solving Eqs.(\ref{GNEE})-(\ref{raman}) in the frequency domain as a set of coupled ordinary differential equations for the resonator modes, with computationally efficient evaluation of the four-wave mixing terms in the time domain via the fast Fourier transform routine\cite{NCME}. This approach permits to easily include in the modeling the frequency dependence of nonlinear terms, such as TPA, as well as of the effective mode area. We used a quantum noise input (one photon per mode), and considered a microresonator with quality factor $Q=\pi n_g/\alpha'\lambda_0=3.5\times10^{5}$ operating in the critical-coupling regime, i.e., we set $\alpha'=\theta=2\alpha_dL$, so that $\theta=0.004$ at $\lambda_0=2.6\:\mu m$. We pumped with $P_{in}=1$ W of input cw power and used $\sigma=3.7\times10^{-21}m^2$, $\mu=4.7$ and $\gamma_{3PA}=0.025\:cm^3/GW^2$ \cite{Lau14}-\cite{TWang13}.
\begin{figure}[htbp]
\centerline{\includegraphics[width=0.8\columnwidth]{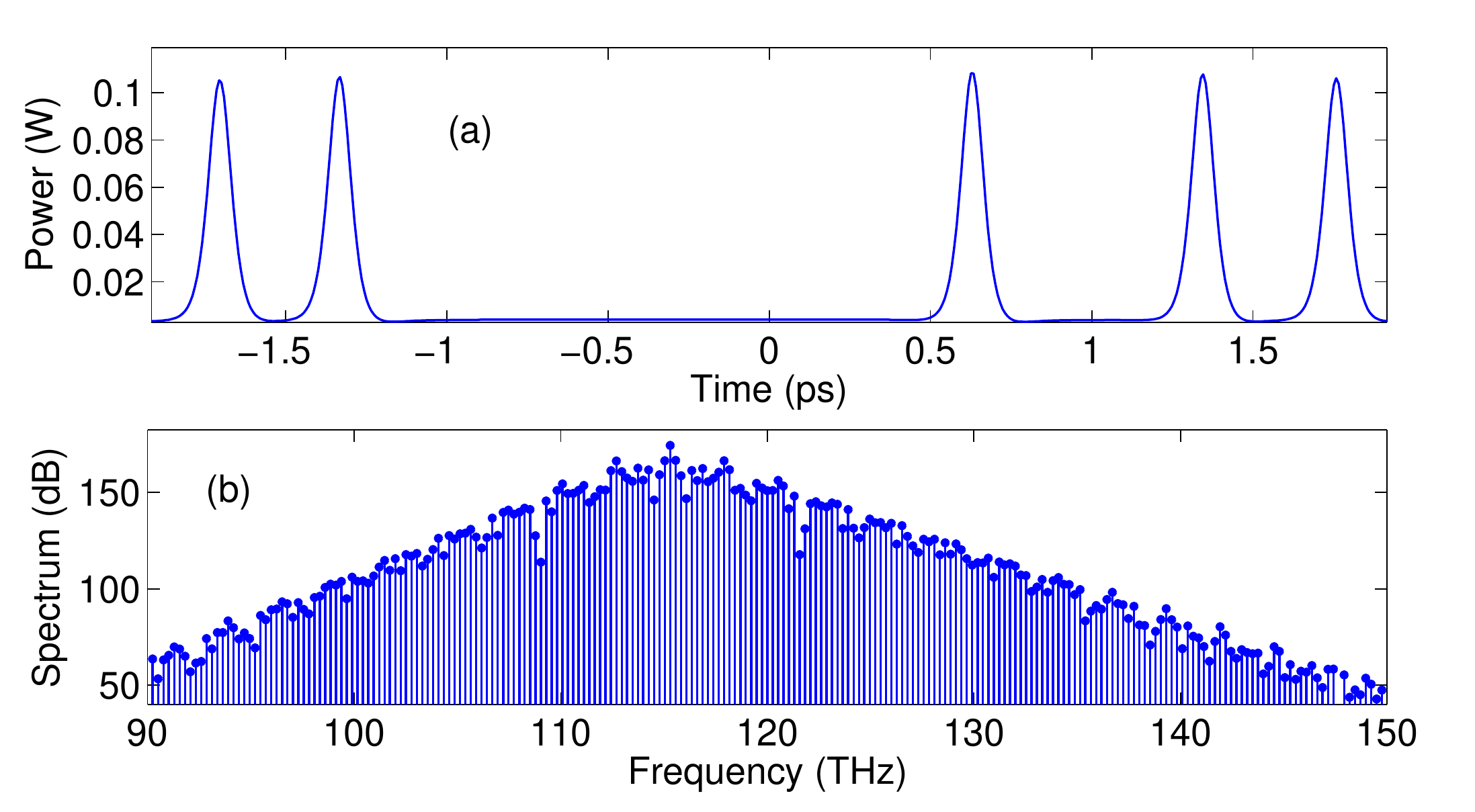}}
\caption{Output soliton frequency comb as in Fig.\ref{fig:fig4}: (a) power and (b) spectral profile. }
\label{fig:fig5}
\end{figure}

Let us consider first the dynamics of frequency comb generation when pumping the TM mode. Fig.\ref{fig:fig4}(a) shows the temporal evolution of the intracavity intensity in the TM mode of the large waveguide of Fig.\ref{fig:fig2}(b) for a pump wavelength $\lambda_0=2.6\:\mu m$ and no cavity detuning, i.e., $\delta_0=0$: the corresponding $t_R=3.83$ ps and the free spectral range (FSR) is equal to 261 GHz. As can be seen, a stable soliton pattern is generated, which is composed of a bound soliton quadruplet and an additional isolated soliton. The corresponding temporal (spectral) profile of the power coupled out of the resonator is shown in Fig.\ref{fig:fig5}a(b). 

\begin{figure}[htbp]
\centerline{\includegraphics[width=0.8\columnwidth]{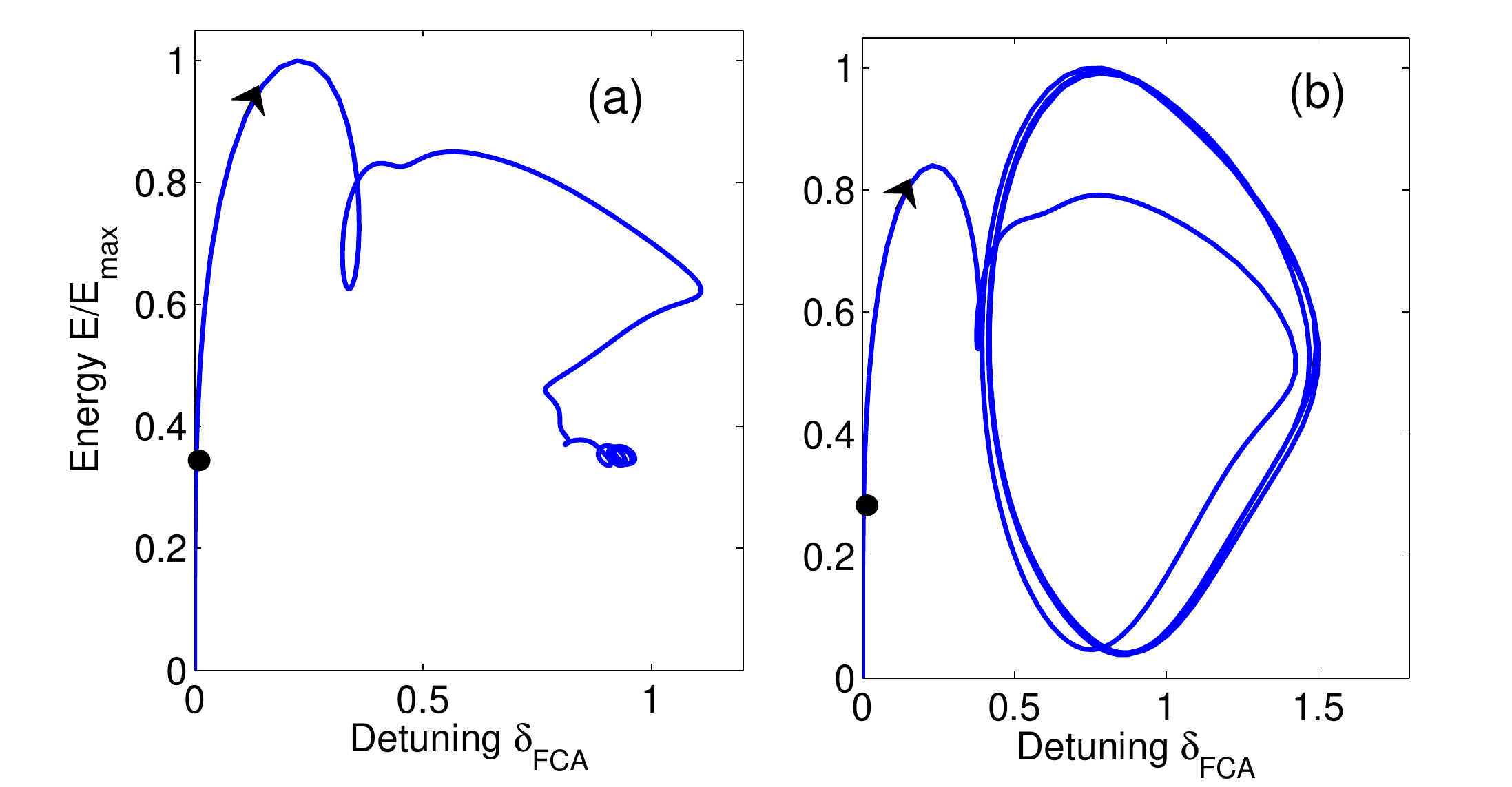}}
\caption{Comb energy vs. FCA-induced detuning: (a) as in Fig.\ref{fig:fig4}; (b) as in Fig.\ref{fig:fig9}. The black dot indicates the initial values and the arrow shows the direction of evolution.}
\label{fig:fig5bis}
\end{figure}

The spectral intensity in Fig.\ref{fig:fig4}(b) shows that a transient primary frequency comb after about 1 ns is followed by the generation of stable and coherent soliton comb. The evolution of the field energy in Fig.\ref{fig:fig4}(c) shows the relaxation towards a stable fixed value associated with the $N=5$ soliton ensemble. 
In order to arrive to a stable multi-soliton comb, it is necessary to properly adjust the FCT (here $\tau_{eff}=320\:ps$). As shown in Refs.\cite{griff14a,griff14b,turner10}, this can be achieved by tuning the reverse bias voltage applied to a PIN structure embedding the microring: $\tau_{eff}=320\:ps$ corresponds to about 2 V bias. 

Here the mechanism for soliton generation is, first, the noise-induced modulation instability of the cw cavity solution, which leads to a periodic pulse pattern. This is followed by the development of nonlinear cavity detuning introduced by FCD, which leads to the generation of isolated cavity solitons. Fig.\ref{fig:fig5bis}(a) shows the $\tau-$evolution towards a stable fixed point of the comb energy and detuning $\delta_{FCD}=\sigma\mu \left\langle N_c\right\rangle L_d/2\approx0.95$, where $L_d\equiv t_0^2/|\beta_2(\lambda_0)|$ and $t_0=65\:fs$. The resulting soliton amplitude may be approximated as $|A(t)|=\eta sech(\rho t/t_0)$ \cite{wab93}, where $\eta=\sqrt{\sigma\mu \left\langle N_c\right\rangle /\gamma}$, $\rho=\sqrt{2\delta_{FCD}}$ and $\gamma=2\pi n_2/\lambda_0$. One predicts for Fig.\ref{fig:fig4}-\ref{fig:fig5} the soliton peak intensity $I_s=\eta^2\simeq 2\:GW/cm^2$ and temporal duration $t_s=1.763 \sqrt{|\beta_2(\lambda_0)|/(\sigma\mu \left\langle N_c\right\rangle )}\simeq 83\:fs$, in excellent agreement with the numerics. Thus FCD leads to cavity soliton generation via the development of a suitable nonlinear cavity detuning.

%
\begin{figure}[htbp]
\centerline{\includegraphics[width=0.8\columnwidth]{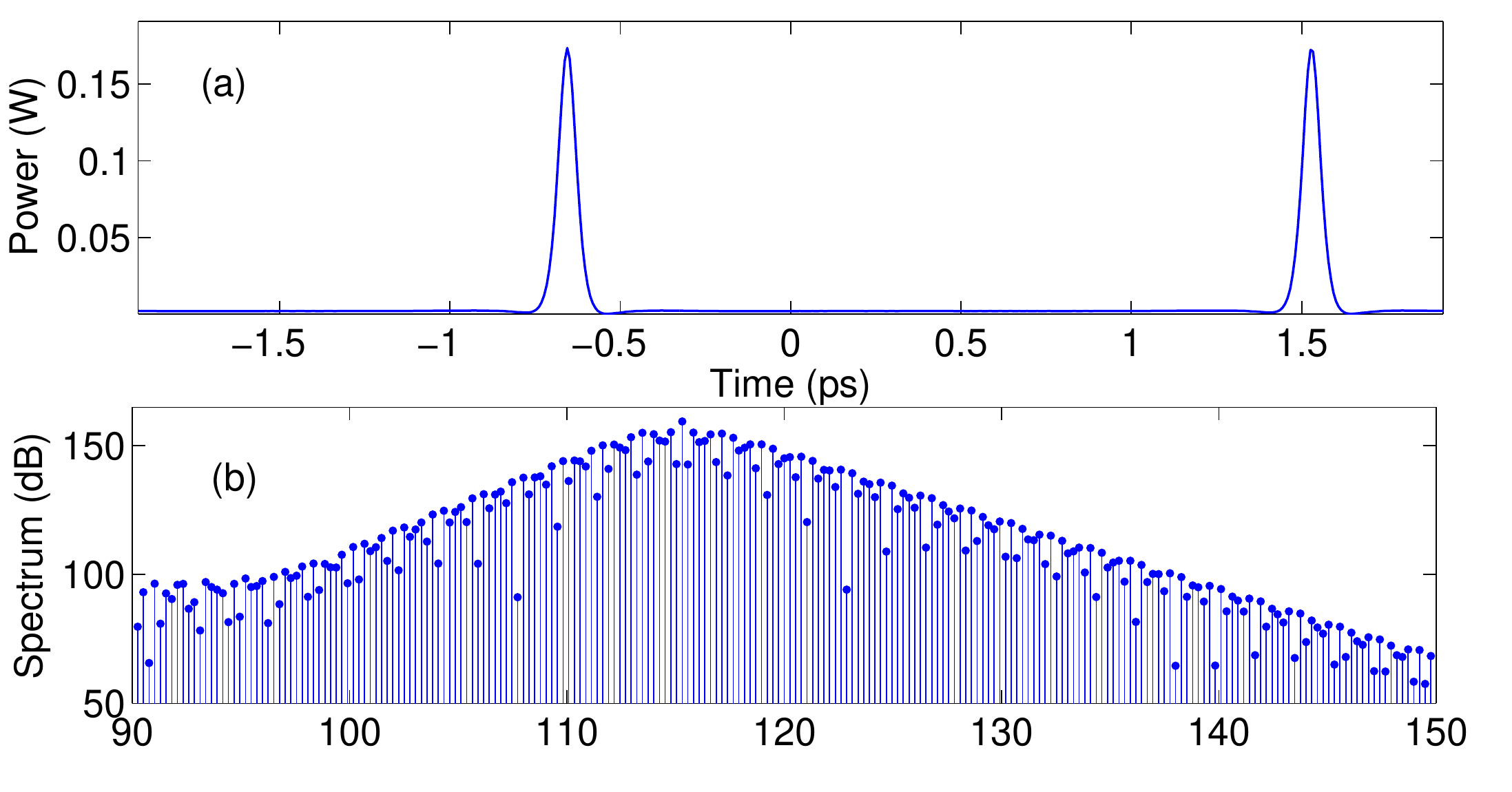}}
\caption{N=2 soliton frequency comb as in Fig.\ref{fig:fig5} with $\bar{\delta_0}=0.05$ and $\tau_{eff}=190\:ps$.}
\label{fig:fig5tris}
\end{figure}
The number of cavity solitons $N$ may be further controlled by introducing a (fixed) nonzero linear cavity detuning: Fig.\ref{fig:fig5tris} shows that a stable $N=2$ soliton comb is obtained with $\bar{\delta_0}=0.05$, so that $\delta_0L_d=0.72$. Correspondingly, the FCD-induced detuning is reduced to $\delta_{FCD}=0.58$.
\begin{figure}[htbp]
\centerline{\includegraphics[width=0.8\columnwidth]{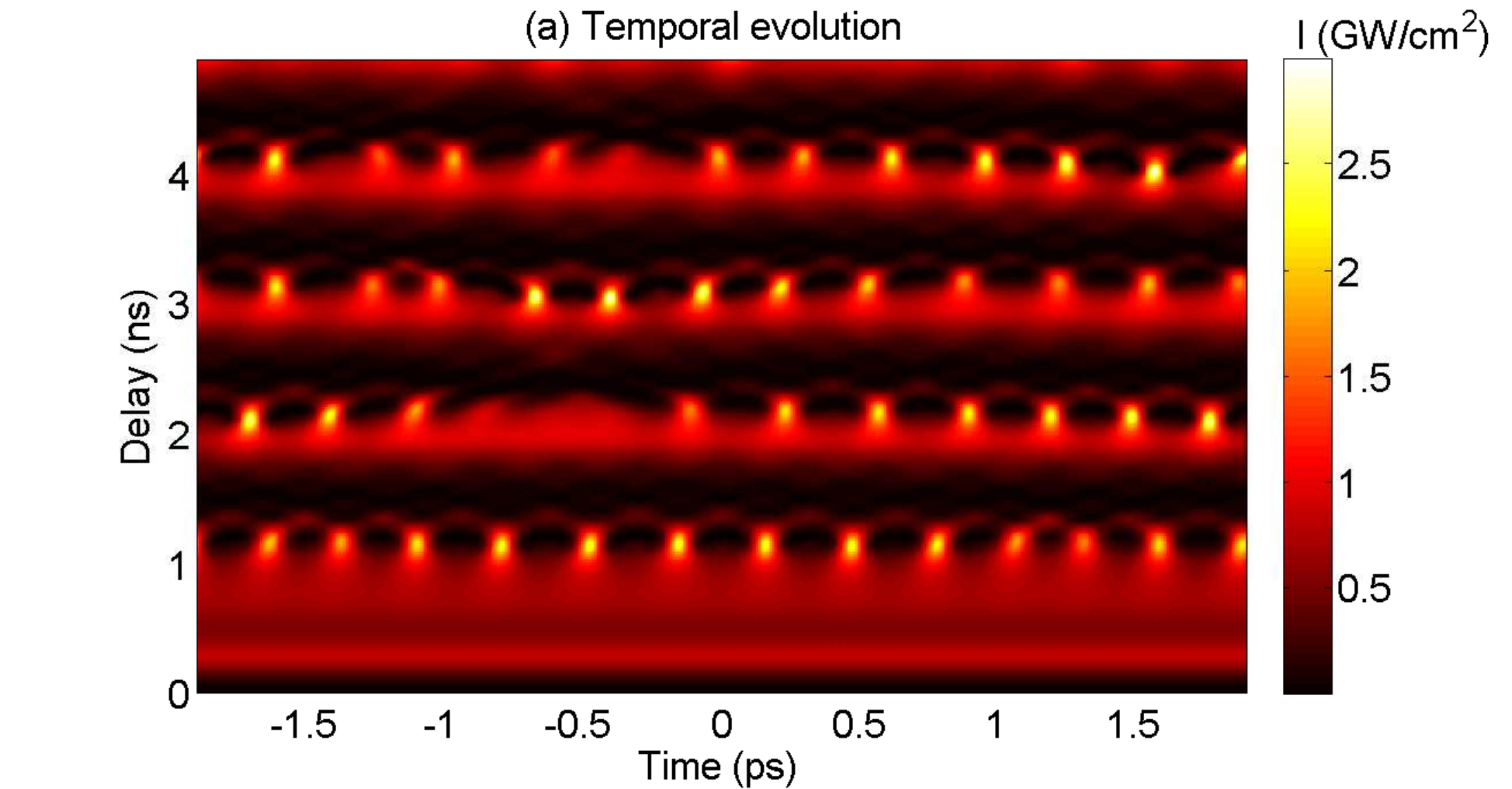}}
\centerline{\includegraphics[width=0.8\columnwidth]{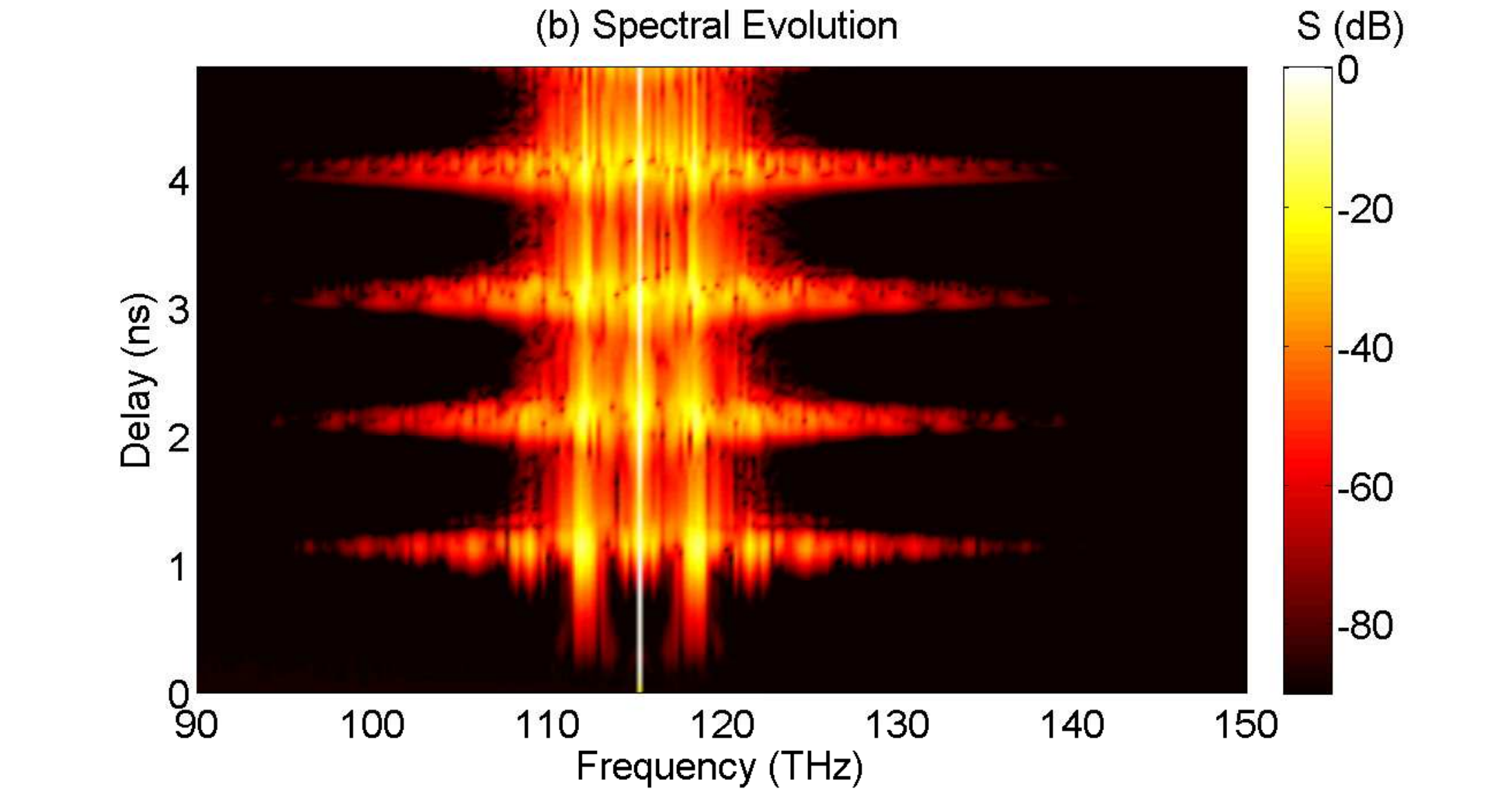}}
\centerline{\includegraphics[width=0.8\columnwidth]{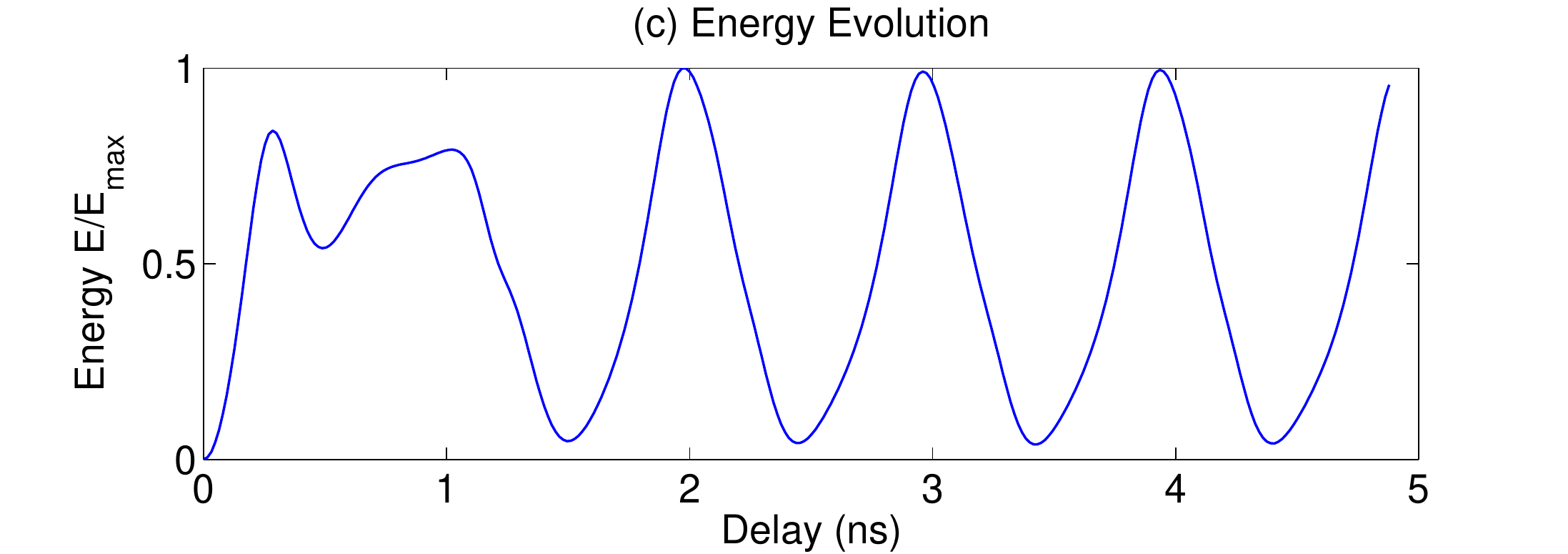}}
  \caption{Same as in Fig.\ref{fig:fig4}, with $\tau_{eff}=400\:ps$.}
\label{fig:fig9}
\end{figure}

For the formation of a stable soliton comb, the FCT should be comprised between a lower and an upper bound. For example, as $\tau_{eff}$ grows larger than a certain critical value, a Hopf bifurcation into an oscillating comb is observed, see Fig.\ref{fig:fig9} where $\tau_{eff}=400\:ps$. In this case the comb exhibits $\tau-$periodic breathing: spectral broadenings in Fig.\ref{fig:fig9}(b) are associated with the emergence of a quasi-periodic pulse train. Moreover Fig.\ref{fig:fig9}(c) and Fig.\ref{fig:fig5bis}(b) show that the intracavity energy exhibits periodic explosions.  

\begin{figure}[htbp]
\centerline{\includegraphics[width=0.8\columnwidth]{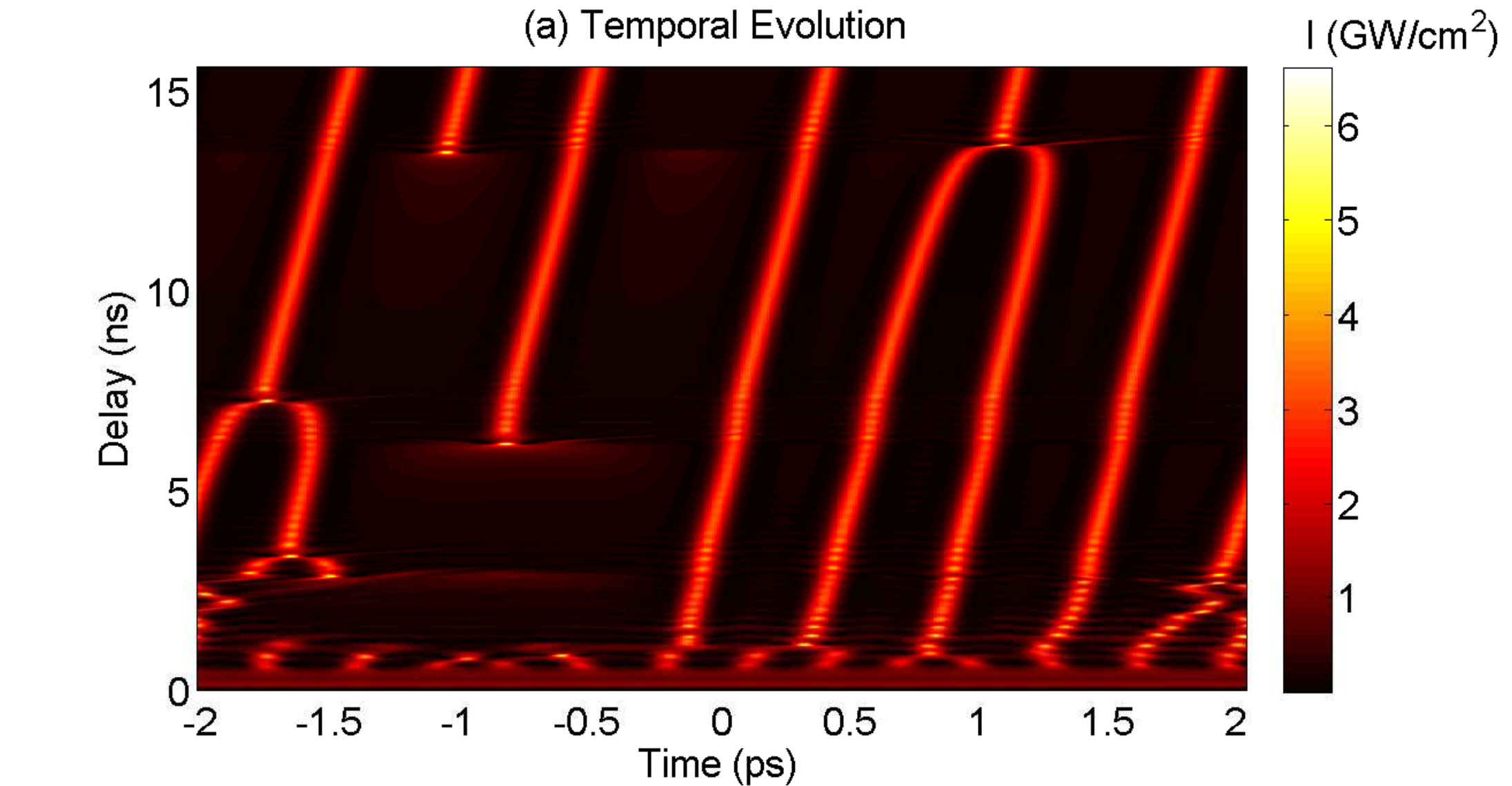}}
\centerline{\includegraphics[width=0.8\columnwidth]{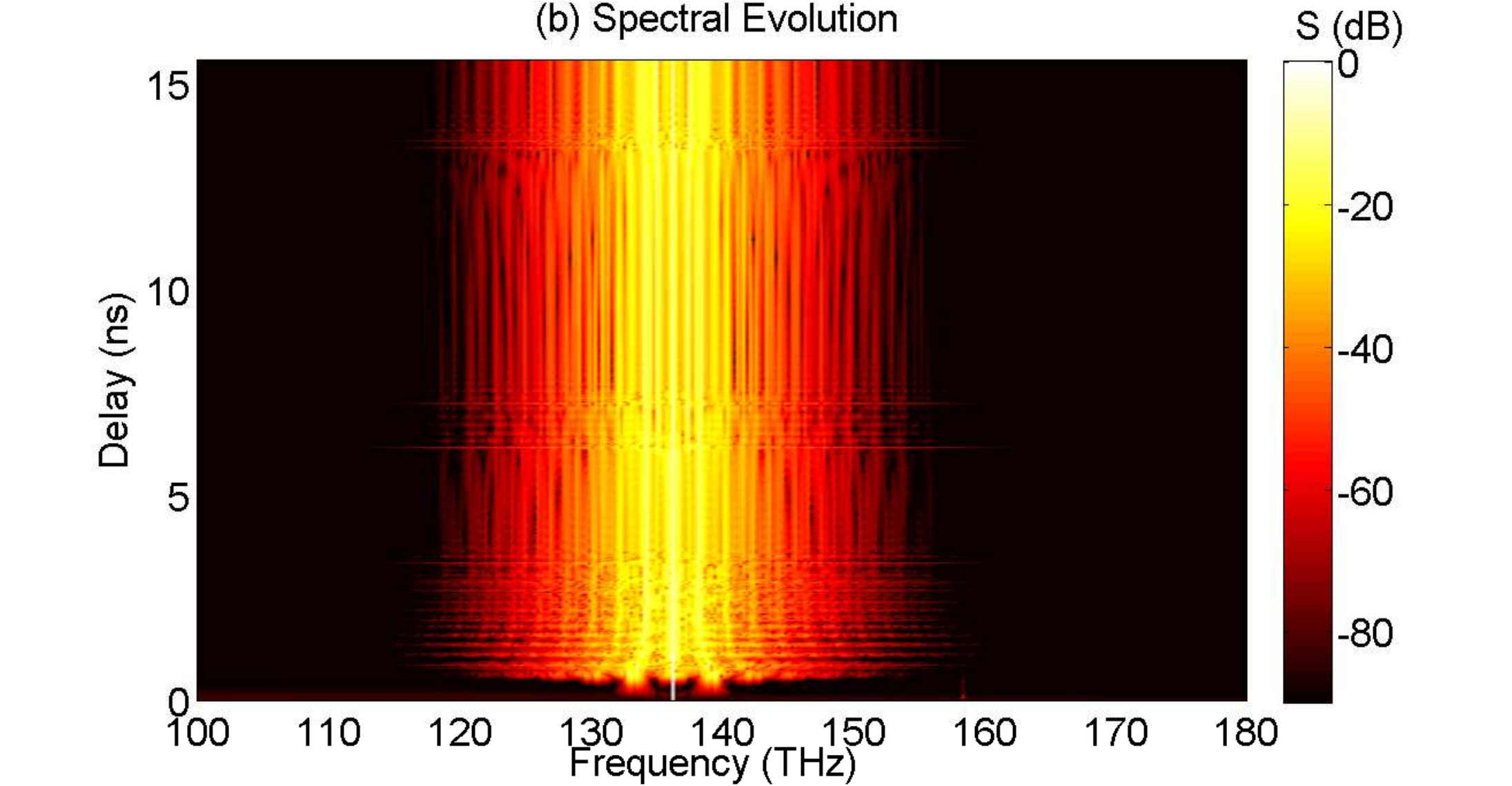}}
\centerline{\includegraphics[width=0.8\columnwidth]{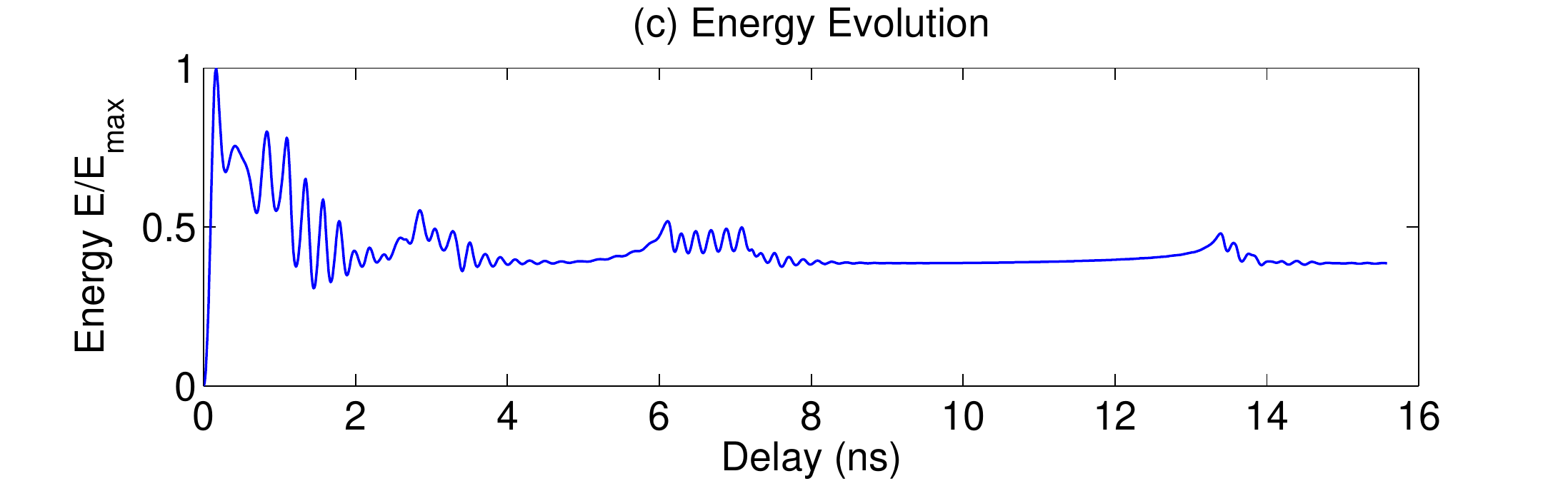}}
  \caption{Soliton frequency comb generation from TM mode of small microring as in Fig.\ref{fig:fig2}(a): (a) temporal evolution; (b) spectral evolution of intracavity intensity; (c) energy evolution. Here $\tau_{eff}=100\:ps$ and $\lambda_0=2.2\:\mu m$}
\label{fig:fig8}
\end{figure}

The described self-induced soliton comb dynamics is also observed for different microring geometries and pump wavelengths. Fig.\ref{fig:fig8}, obtained when pumping at $\lambda_0=2.2\:\mu m$ the TM mode of the small microring of Fig.\ref{fig:fig2}(a), shows that the long-term behavior of a $N=6$ soliton comb is subject to strongly dissipative dynamics: soliton annihilation (upon collision) is accompanied by simultaneous soliton creation at a different temporal position, so that the total soliton number is conserved, to keep the FCD-induced detuning unchanged. Fig.\ref{fig:fig8}(b,c) show that, in correspondence with each soliton annihilation/creation event, a transient spectral broadening occurs, accompanied by a small peak in the energy evolution. 
  
%
%

\begin{figure}[htbp]
\centerline{\includegraphics[width=0.8\columnwidth]{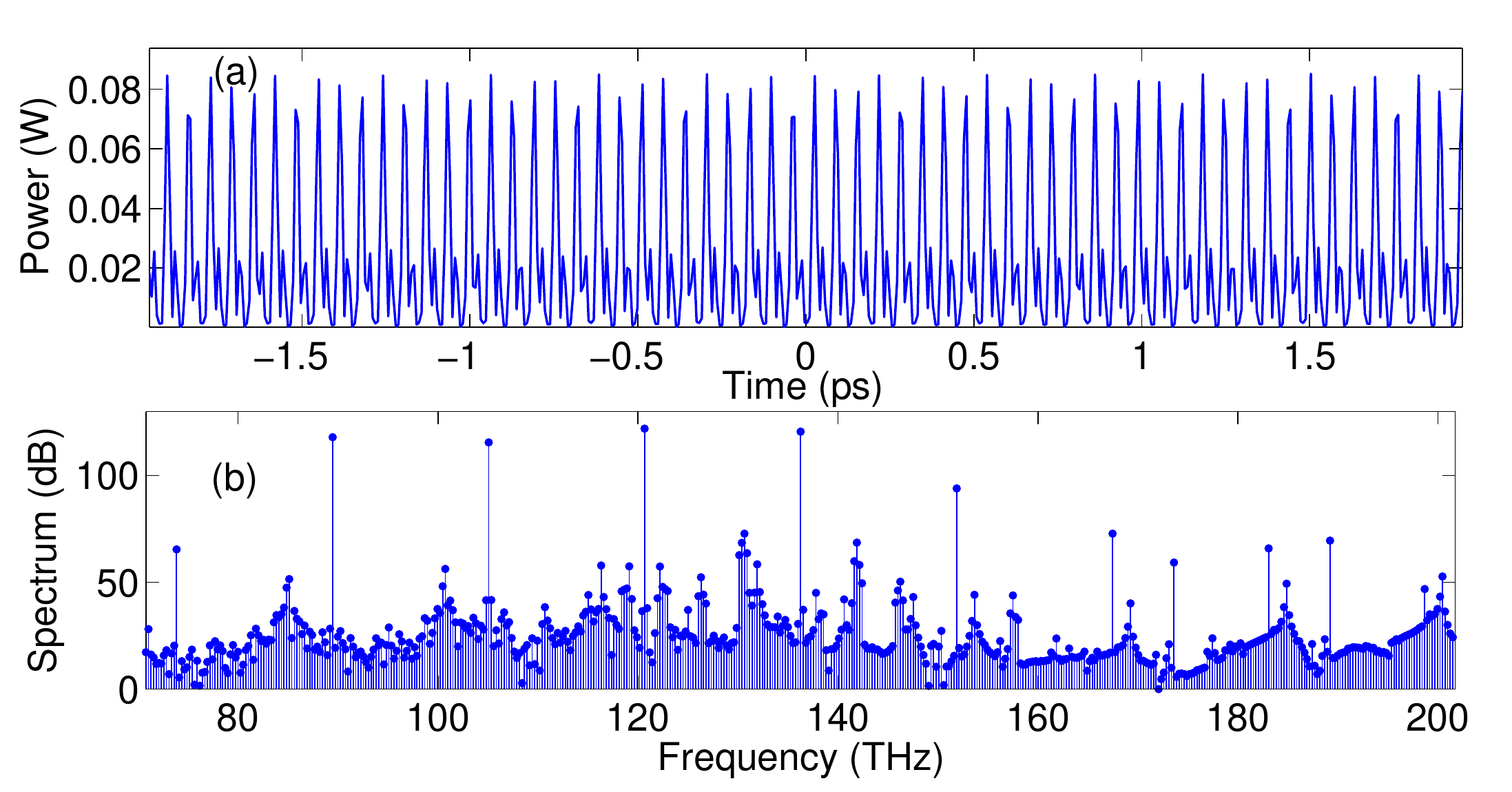}}
\caption{Raman frequency comb from TE mode of large microrings as in Fig.\ref{fig:fig2}. Here $\tau_{eff}=100\:ps$ and $\lambda_0=2.2\:\mu m$; (a) power and (b) spectral profile. }
\label{fig:fig7}
\end{figure}

When pumping the TE mode of the microring, the presence of parallel Raman gain dramatically changes the nature of the generated frequency comb. To ensure that the peak of Raman gain at 15.6 THz shift from the pump occurs for an integer multiple (e.g., $M=61$) of the cavity FSR, we slightly increased the ring diameter (from $50\:\mu m$ to $51.3\:\mu m$). Fig.\ref{fig:fig7}(b) shows that a Raman frequency comb is generated, including three cascaded Raman Stokes lines of nearly equal intensity as the pump wave, as well as three anti-Stokes comb lines and a weaker fourth Stokes line at about 70 THz, for an octave-spanning Raman comb bandwidth in excess of 100 THz.   
Fig.\ref{fig:fig7}(a) shows that the Raman comb results into the generation of a pulse train with about 15 fs duration, which under suitable spectral post-processing could be compressed down to the single-cycle regime. Because of the large Raman gain ($10^4$ times the silica value) in silicon, SRS remains the main comb generation mechanism also if the Raman gain shift is not precisely equal to a multiple of the FSR.    

In conclusion, we demonstrated that a suitable control of the FCT may enable a new route for the generation of stable coherent soliton frequency combs, and predicted the generation of octave-spanning Raman frequency combs in the MIR by using silicon microresonators.

This research was funded by Fondazione Cariplo (grant no. 2011-0395),
the Italian Ministry of University and Research
(grant no. 2012BFNWZ2), and the Swedish Research Council (grant no. 2013-7508).

\end{document}